\documentclass[preprint2]{aastex}

\shorttitle{Extended SZ map of RXJ137-1145}
\shortauthors{Pointecouteau et al.}

\begin{document}

\def\etal{et al. }
\def\araa{{\it Ann.\ Rev.\ Astron.\ Ap.}}
\def\aplet{{\it Ap.\ Letters}}
\def\aj{{\it Astron.\ J.}}
\def\apj{ApJ}
\def\apjl{{\it ApJ\ (Lett.)}}
\def\apjs{{\it ApJ\ Suppl.}}
\def\aas{{\it Astron.\ Astrophys.\ Suppl.}}
\def\aa{{\it A\&A}}
\def\mnras{{\it MNRAS}}
\def\nature{{\it Nature}}
\def\pasa{{\it Proc.\ Astr.\ Soc.\ Aust.}}
\def\pasp{{\it P.\ A.\ S.\ P.}}
\def\pasj{{\it PASJ}}
\def\pre{{\it Preprint}}
\def\aph{{\it Astro-ph}}
\def\sovlet{{\it Sov. Astron. Lett.}}
\def\adspr{{\it Adv. Space. Res.}}
\def\expas{{\it Experimental Astron.}}
\def\ssr{{\it Space Sci. Rev.}}
\def\inpress{in press.}
\def\souspresse{sous presse.}
\def\inprep{in preparation.}
\def\enprep{en pr\'eparation.}
\def\submit{submitted.}
\def\soumis{soumis.}

\def\ergs{ergs s$^{-1}$}
\def\ergscm{ergs s$^{-1}$ cm$^{-2}$}
\def\pho{ photons cm$^{-2}$ s$^{-1}$ }
\def\phokev{ photons cm$^{-2}$ s$^{-1}$ keV$^{-1}$}
\def\ap{$\approx$ }
\def\ep{$\rm e^\pm$ }
\def\mjyb{mJy/beam }

\def\vp{{ v_{p}}}
\def\te{{T_{e}}}
\def\rc{{r_{c}}}
\def\yc{{y_{c}}}
\def\neu{{n_{e}}}

\def\hzero{{H_{0}}}
\def\qzero{{q_{0}}}
\def\tcmb{{T_{cmb}}}
\def\cd{{C_{dust}}}
\def\td{{T_{dust}}}
\def\nd{{n_{dust}}}
\def\inu{{I_{\nu}}}
\def\fnu{{F_{\nu}}}
\def\bnu{{B_{\nu}}}
\def\mecdeux{{m_{e}c^{2}}}
\def\msol{{M$_{\odot}$}}

\newcommand{\hms}[3]{#1$^h$#2$^m$#3$^s$}
\newcommand{\dms}[3]{#1$^\circ$#2$^m$#3$^s$}

\title{Extended SZ map of the most luminous X-ray cluster, RXJ1347-1145} 

\author{E. Pointecouteau, M. Giard \\
Centre d'Etude Spatiale des Rayonnements \\ 
9 av du colonel Roche, BP4346, F-31028 Toulouse cedex 4, France \\
pointeco@cesr.fr}

\author{A. Benoit \\
Centre de Recherche des Tr\`es Basses Temp\'eratures \\
25 avenue des Martyrs, BP 166 , F-38042, Grenoble Cedex 9, France }

\author{F.X. D\'esert \\
Laboratoire d'Astrophysique de l'Observatoire de Grenoble \\
414 rue de la piscine, F-38041 Grenoble Cedex 9, France }

\author{J.P. Bernard, N. Coron, J.M. Lamarre \\
Institut d'Astrophysique Spatiale \\ 
B\^at 121, Universit\'e Paris-Sud, F-91405 Orsay cedex, France}


\begin{abstract}
We present in this letter a high resolution (22'' FWHM) extended map  at 2.1mm
of the Sunyaev-Zel'dovich effect toward the most luminous X-ray cluster,
RXJ1347-1145. 
These observations have been performed with the DIABOLO photometer
working at the focus of the 30m IRAM radiotelescope.
We have derived a projected gas mass of $(1.1 \pm 0.1)\times 10^{14}\,
h_{50}^{-5/2}$M$_{\odot}$ within an angular radius of $\theta=74''$ (ie:
projected  radius of $0.6$Mpc, $H_{0}=50$km/s/Mpc, $\Omega_m=0.3$,
$\Omega_\Lambda=0.7$).  
This result matches very well the expected gas mass from the cluster models of
X-ray data. With an unprecedented sensitivity level our measurement does not 
show significant departure from a spherical distribution.
The data analysis also allows us to characterize the 2.1mm
flux of a well known radio source lying in the center of the cluster:
$F_{RS}(2.1\textrm{mm})=5.7\pm 1.6$mJy.

\end{abstract}


\section{Introduction\label{intro}}

The statistical properties of galaxy clusters (shape, structures, size,
temperature, mass) depend  strongly on the geometry of the Universe. Their
study provides some robust constraints on cosmological models, structure
formations and evolution \citep{oukbir97,sadat98,bahcall98}. 
This strong coupling makes the study of massive and distant clusters a very
useful tool for cosmology.
The intergalactic gas component can be observed through its Bremsstrahlung
emission at X-ray wavelengths. It can also be detected from submillimeter to
radio wavelengths via the Sunyaev-Zel'dovich effect \citep{sunyaev72}. 
Whereas the X-ray emission depends on the square of the gas density, the
Sunyaev-Zel'dovich (SZ hereafter)
effect is linearly dependent on this quantity. For this reason, the SZ effect
is proportional to the column density and thus to the line of sight 
integrated gas mass.
Moreover the SZ signal does not suffer from the
brightness decrease of the radiations (in particular
the X-ray one) due to the expansion of the Universe  (see a review by
\citet{birkinshaw99}).  

The RXJ1347-1145 cluster is known as the most luminous X-ray cluster to
date. It has been detected in the ROSAT All Sky Survey and furthermore studied
with the ROSAT-HRI and the ASCA-GIS2 instruments
\citep{schindler95,schindler97}. With an intrinsic bolometric X-ray
luminosity of $L_{bol}=21\times 10^{45} \, h_{50}^{-2}$\ergs, it shows a very 
peaked X-ray emission profile with an angular core radius of $\theta_c=8.4\pm
1.8$''. It also presents a very strong cooling flow in its central region, with
a corresponding accreting rate of
$\dot{\textrm{M}}_{cool}=3000$M$_{\odot}$/yr. It is a distant cluster with
$z=0.45$, so that the angular distance is $1670\, h_{50}^{-1}$Mpc
(490pc/arcmin, with $H_0=50$km/s/Mpc, $\Omega_m=0.3$, $\Omega_\Lambda=0.7$; 
410pc/arcmin in a standard CDM model). The value of the core radius with these 
two sets of cosmological parameters is respectively equal to 68kpc 
($\Lambda$CMD model) and to 57kpc (standard CDM model).  
The total binding mass derived from the X-ray data within 1Mpc is
$M_{tot}=5.8\times 10^{14}\, h_{50}^{-1}$M$_{\odot}$. This value has to be 
compared to those obtained from the optical gravitational lensing follow-up 
achieved by \citet{fischer97} and \citet{sahu98}. Within the same radius, they 
provide a value of  $M_{tot}=1.7\times 10^{15}\, h_{50}^{-1}$M$_{\odot}$ 
\citep{fischer97}.

In a previous paper, we have reported the detection of a very  strong SZ effect
in the direction of RXJ1347-1145 with the DIABOLO photometer
\citep{pointecouteau99}.  We have presented a map of the cluster
central region  ($2' \times 1'$). The corresponding
Comptonization parameter was $y(0)=12.7^{+2.9}_{-3.1}\times 10^{-4}$ 
(1$\sigma$ error bars). 
During the 1999 DIABOLO run, we performed an extended mapping of
RXJ1347-1145  over a 4' by 4' field.
The size of the first map was rather small so that only 1D average slice was
used to compare with the X-ray data.
Here the size of the map allows to make full use of 2D information on the
cluster. Moreover, the computation of a cluster radial profile from a 2D map
allows to discard most of the effects induced by point sources.

In this paper, we present the resulting extended map of this cluster at
2.1mm. In a first part, we detail the observations
procedures and characteristics (see Sec~\ref{obs}). In a second
time (see Sec~\ref{red}), we describe the data processing. In
section~\ref{resu}, we analyze the astrophysical data. In the last part (see
Sec~\ref{conclu}), we discuss the results. 
Throughout this paper, we will use the following values for the cosmological
parameters: $H_0=50$km/s/Mpc, $\Omega_m=0.3$, $\Omega_\Lambda=0.7$.

\section{Observations\label{obs}}

The DIABOLO instrument is a dual channel photometer working at 1.2 and 2.1mm.
The detectors are bolometers cooled down to 0.1K using an open cycle
$^{4}$He-$^{3}$He dilution refrigerator \citep{benoit00}. Two
thermometers associated to a heater and a PID digital control system are
used to regulate the temperature of the 0.1K plate.
There are three adjacent bolometers per channel, arranged in an
equilateral triangle at the focus of the telescope. For a given channel,
each bolometer is coaligned with one bolometer of the second channel, both
looking toward the same sky direction.
For the observations presented here, DIABOLO was installed at the focus of the
IRAM 30 meter radio telescope at Pico Veleta (Spain). This configuration
allows us to achieve a 22'' resolution at 2.1mm. 
The 30 meter telescope focus being of Nasmyth type, the rotation of the field
has to be taken into account in the sky maps reconstruction.
\citet{desert98} have described the experimental setup and
reported the first DIABOLO SZ detections.  

RXJ1347-1145 has been observed in January 1999 for a total integration time
of 17 hours. The X-ray emission center:
$\alpha_{2000}=13^{\rm{h}}47^{\rm{m}}31^{\rm{s}}$,
$\delta_{2000}=-11^{\circ}45'11"$ \citep{schindler97}, has been taken as the
map center. 
Each sequence of observation has been performed in the right ascension
coordinates, using the Earth rotation as right ascension drift such that the
telescope is kept fixed in local coordinates. 
The basic observation sequence was a $240''\times 240''$ map in right
ascension-declination coordinates, with a 10'' declination steps.
The wobbling secondary mirror of the 30m IRAM telescope has been used at a
frequency of 1Hz and with a modulation amplitude of 150''.  
The wobbling is horizontal (eg: at constant elevation), thus not aligned with
the scanning direction. In order
to remove systematic signal drifts that are produced by the antenna
environment, we used  alternatively the positive and negative beam to map the cluster. 
We have produced 89 such maps.

\section{Data reduction and calibration\label{red}}

The data processing includes all the different steps described by
\citet{pointecouteau99}. This includes a correction from the 
cosmic-rays impacts, a synchronous demodulation of the signal, a subtraction
of the atmospheric signal correlated between the two channels, a correction
from the atmosphere opacity, a reprojection of each single observation on a
final RA-DEC grid, a baseline subtraction supported by the edges of the map and
the computation of a final map at 1.2 and 2.1mm by coadding the 3
bolometers of each channel.
The beam modulation is performed with an amplitude of 150'', so that some of
the ``OFF'' positions lie within the limit of our map. Consequently, the beam
switching has to be taken into account in the data analysis (see
Sec.~\ref{szsignal}). 

Throughout the run, the pointing verifications have been performed in the
direction of QSOs lying at about the same declination as the cluster.  
The planet Mars has been used as a calibration target and to map the beam
pattern. Assuming Mars is a point source with respect to the DIABOLO's beam
(angular diameter around 5'' in January 1999), the accuracy in the absolute
calibration is better than 20\% at 1.2mm and 15\% at 2.1mm. 
Mars has directly been mapped in right ascension-declination during this run.
This method provides a direct view of the beam shape including eventual
systematic effect as the beam elongation in the right-ascension direction due
to the conjugated effects of the scanning drift speed and the bolometer time
constant.

\section{Results and data analysis \label{resu}} 

The reduced map for RXJ1347-1145 at 2.1mm is shown on Fig.~\ref{fig1}a.
It has been smoothed using a Gaussian filter with 20'' FWHM. The
contours overplotted correspond to 1, 2, 3, 4 and 5 $\sigma$ detection levels
(negative  and positive contours are respectively drawn with solid and dashed
lines). The noise level on this map is 1mJy/beam in a $10''\times10''$ pixel
(equivalent to 0.5mJy/beam in the 22'' DIABOLO beam and to 0.3mJy/beam
(equivalent to $y=3\times 10^{-5}$) in the 
30'' FWHM effective beam after smoothing). This map exhibits a strong  and
extended SZ decrement in the direction of the cluster. 
Unfortunately, the DIABOLO 1.2mm data are very noisy. No positive SZ signal
could be extract from them. We used them to subtract most of the atmospheric
emission from the 2.1mm data.

The signal does not seem to follow a symmetric circular distribution as
expected from the X-ray data and from the commonly  used $\beta$-model.
The first look at the map argues in favor of substructure  in the signal, and
so  far, in the gas distribution. 
The X-ray map has been overplotted on the SZ map (see Fig.~\ref{fig1}c). It
has been computed from the whole ROSAT/HRI observations (35ks 
exposure time) obtained from the ROSAT database (the map published by
\citet{schindler97} included only a third of those data).

As the SZ signal is linearly dependent of the gas density (scaled with $n_e$),
the X-ray emission is dominated by contributions from regions of higher
density (scaled with $n_e^2$). Furthermore, the X-ray and SZ signals are not 
sensitive to the same part of the gas. 
What appears as differences between the SZ and the X-ray spatial
distributions could be associated to different gas phases with various physical
states (density, pressure, temperature). In the following, we will compare in
detail the measured SZ map to the expected SZ distribution deduced from the
symmetrical cluster X-ray model.

\subsection{Modeling the SZ signal \label{szsignal}}

From the ROSAT-HRI X-ray map and the ASCA X-ray spectrum of RXJ1347-1145 
\citet{schindler95,schindler97} have  
extracted the cluster physical parameters: $T_e=9.3\pm 0.5$keV, 
$n_{e0}=9.4\times 10^{-3}$cm$^{-3}$, $\theta_{c}=8.4\pm 1.8$'' and
$\beta=0.56\pm 0.04$ ($1\sigma$ errors), assuming a $\beta$-model for the gas
distribution \citep{cavaliere76}. Using those parameters, we can compute the
Comptonization parameter expected toward the cluster center: 
$y_{exp}(0)=(8.4\pm 2.7)\times 10^{-4}\, h_{50}^{1/2}$.
Following this approach, we used a $\beta$-model to
describe  our SZ signal. Because of the circular symmetry induced  on the
projected sky by this kind of model, we chose to test different
$\beta$-models on the radial profile resulting from our 2.1mm map (see
Fig.~\ref{fig2}). 
To model the SZ signal, we have taken into account the integration of the SZ
spectrum on the DIABOLO passbands and the integration of the gas distribution
over the DIABOLO beam shape.
To perform a rigorous fit
of the DIABOLO data, we need to take into account in  
the model every step of the observing procedure and of the data reduction
procedure . So whatever the model we used,  we considered it as a real sky
and we reproduced on it all the DIABOLO's observations performed in the
direction of RXJ1347-1147. 
(This includes the wobbling positions and amplitudes.). The
resulting simulated data set of observations has been reduced in the same way
as the real dataset by following all steps of the DIABOLO pipeline
(demodulation, base line subtraction, reprojection,... see Sec~\ref{red}). 
 
The resulting model for the flux of the SZ decrement per beam can be expressed
as follow:

\begin{equation}
F(\bar{\nu},\overrightarrow{\Omega})=y(0)\int{\tau{\nu}SZ(\nu,T_g)d\nu}\int{P(\overrightarrow{\Omega})L(\overrightarrow{\Omega}-\overrightarrow{\Omega'})d\overrightarrow{\Omega'}}
\label{eqsz}
\end{equation}

where $y(0)=(k_B T_g/\mecdeux) \sigma_T \int{n_e(l)dl}$, is the Comptonization
parameter towards the cluster center ($k_B$ is the Boltzmann constant, $m_e$
the electron mass and $c$ the speed of light. $\sigma_T$ is the Thomson cross
section, $n_e(l)$ is the electronic density along the line of sight). 
$\tau(\nu)$ is the normalized DIABOLO band spectral
efficiency. $SZ(\nu,T_g)$ represents the SZ spectrum. It is a numerical
function   of $\nu$ and $T_g$ with a weak dependence on $T_g$ that takes into
account the relativistic corrections \citep{pointecouteau98}.
$P(\overrightarrow{\Omega})$ represents the normalized gas profile projected on
the sky and $L(\overrightarrow{\Omega})$ is the normalized DIABOLO beam shape. 

The cluster central region contains a radio point source, known from the NRAO
VLA Sky Survey (NVSS) \citep{condon98}, which coordinates are
$\alpha_{2000}=$\hms{13}{47}{30.7} and $\delta_{2000}=$\dms{-11}{45}{8.6}
(within the 3'' of the X-ray center). To avoid any contamination and bias of 
the SZ signal due to a residual millimeter emission of this radio point
source, we chose to include its contribution in our model. We considered
this source as a point source with respect to the DIABOLO's beam. This
approach follows the method used by \citet{pointecouteau99}:

\begin{equation}
F(\bar{\nu},\overrightarrow{\Omega})=F_{RS}(\bar{\nu})
L(\overrightarrow{\Omega})  
\label{eqrs}
\end{equation}

$F_{RS}(\bar{\nu})$ is the flux of the point source. 

We have tested a total of five  models on our data. 
The models $A$ and $B$ just include an SZ component. 
In model $A$, we choose the Comptonization parameter, the core radius and
$\beta$  as free parameters. In model $B$, we just let free $y$ and
$r_c$ and fixed $\beta$ to 0.56 (the X-ray value). 
The three other models are a combination of an SZ component and a point
source component.  
The parameters in model $C$ are $y$, $r_c$, $\beta$ and $F_{RS}$.
In model $D$, we let free the $y$, $r_c$ and $F_{RS}$ parameters
($\beta=0.56$).  Finally, for the $E$ model, we have fixed the core radius and
the $\beta$ parameters to their respective X-ray value of 8.4'' and 0.56. We
used the $y$ and $F_{RS}$ as free parameters.  
The models have been tested using a maximum likelihood analysis method. 
The errors are obtained through the integration of the likelyhood function over
the parameter space. The best fit parameters are gathered for the different
cases in table~\ref{table1}. 

As shown by the results from models $A$ and $C$, we can not constraint the core
radius and the $\beta$ parameters simultaneously, because those two parameters
are too strongly coupled. 
For this reason $\beta$ has been fixed to the X-ray value of 0.56 in
models $B$ and $D$. Unfortunately, in those two cases the core radius can also
not be constraint precisely. 
Its determination suffers from a degeneracy and despite of the extent of our SZ
map, we are not able to discriminate between a small and a high core radius
model.  
This is due to the way we have defined the zero level: a baseline
subtraction in the RA direction fitted to 60\% of the data point per line
30\% on each edge). This operation is needed to eliminate the low-frequency
detector noises. Consequently, whatever the model is, the cluster extension is
cut off by the baseline subtraction and the signal to noise ratio does not
allow to make the difference between small and large core radius.
For this reason, we have chosen to use the core radius value derived from
the X-ray analysis, $\theta_c=8.4$'',  to describe the spatial gas
distribution with a $\beta$-model.   
Following this hypothesis, we have adopted the model $E$ as the best fit model. 
The best fit parameters are $y(0)=(7.9 \pm 0.5)\times 10^{-4}$ and
$F_{RS}=5.7\pm1.6$mJy (with 68\% confidence level error bars).
In order to perform an accurate error analysis, we have also propagated the
uncertainty on the X-ray parameter determination through the whole pipeline of
the data analysis. The extra 1$\sigma$ errors are $\Delta y=1.1\times 10^{-4}$
and $\Delta F_{RS}=1.8$mJy.
Because  we just use the information on the X-ray temperature to fix the exact
shape of the SZ spectra (a second order effect), the extra errors induced by
the temperature uncertainty is marginally significant. The most important
difference is driven by the core radius and by the $\beta$ parameters
uncertainties.  
Figure~\ref{fig2} presents the best fit radial profile model (SZ plus point
source signals) overlying the cluster radial profile as seen by DIABOLO at
2.1mm. The residual signal is overplotted as a dashed line.

One can also perform a fit with $y$ as a single free parameter. To do that,
the central part of the map ($\theta_c<30$'') is excluded from the radial
profile computation. The best fit parameter in this case
is $y(0)=(7.8\pm0.4)\times 10^{-4}$. Afterward, the point source
flux can  be derived from the residual map: $F_{RS}=5.3\pm1.0$mJy. 
Those two values are fully  consistent with the previous ones.
(The propagation of the X-ray parameters uncertainties reach to respective
extra errors of $\Delta y=1.0\times 10^{-4}$ and $\Delta \fnu=1.2$mJy).
Our $y$ value is in very good agreement within a 68\% confidence level with the
value expected from X-ray data. Moreover, it is in very good agreement with the
value determined by \citet{komatsu00} from their 21GHz data:
$y(0)=(7.7\pm1.6)\times 10^{-4}$. Finally, within the 2$\sigma$
of our previous determination: $y(0)=12.7^{+2.9}_{-3.1} \times
10^{-4}$ \citep{pointecouteau99}. 
Combining the $y(0)$ value expected from the X-ray data
($y_{exp}(0)=(8.4\pm 2.7)\times 10^{-4}\, h_{50}^{1/2}$) and the one we
estimated from our SZ measurements, we are able to derived the $H_0$ value:
$H_0=44\pm6$km/s/Mpc. 
The error quoted on $H_0$ just includes the uncertainty
on our $y(0)$ determination. If the uncertainties on the X-ray parameters are
taken into account, the error on $H_0$ become $\pm 15$km/s/Mpc.
Obviously this does not take into account any other uncertainty or any
systematic errors due to the cluster geometry or to the hypothesis concerning
the gas isothermality or the hydrostatic equilibrium.

The residual signal resulting from the data map and the best fit model map
difference is presented on figure~\ref{fig1}b (A Gaussian 20'' FWHM filtering
has been performed). The 1, 2 and 3 $\sigma$ detection levels have been
overplotted (solid contours for the negative signal). This map is mostly
compatible with noise, the higher deviation being a decrement ($-3.1\pm0.9$mJy)
located to the North-East of the cluster center ($\Delta \alpha
=+35''$,$\Delta\delta=+30''$), with a  significance: 3.5$\sigma$. 

\citet{komatsu00} have mapped RXJ1347-1145 at at 150GHz with the
NOBA/NRO instrument. They  get a $2'\times2'$ map of the cluster center.
To compare our work to theirs, we have divided the
center of our map in the same 4 regions ($50''\times50''$ square), SE, NE, NW
and SW, as they did. We have integrated the flux in each
region. The value derived for each region can then be compared to the one
extracted from the NOBA/NRO map. The results are gathered in
table~\ref{table2}. We also show the integrated values for our residual map.  
We do not confirm the negative excess for the SE quadrant. Instead, our
measurement is symmetrical within the noise level. If any departure is to be
searched in our map, then it is a positive excess in the SW quadrant:
$2.8\pm1.4$. Despite the low significance level of this excess and refering to
our residual map, we can suggest that this excess is compatible with the
presence of a second point source in this region.
This hypothesis is supported by the point source detected at
350GHz with SCUBA and at 8.46GHz with the VLA by Komatsu et al (private
communication). Its VLA position is $\alpha_{2000}$=\hms{13}{47}{27.72},
$\delta_{2000}$=\dms{-11}{45}{52.86}. The SCUBA and VLA respective fluxes are
$17.9\pm 4.8$mJy and $0.490 \pm 0.043$mJy. 
This source seems to be an infrared source. A millimeter
residual, provided by the millimeter tail of a dust emission, could contribute
to our signal. Complementary millimeter observations are needed for this
source.

\subsection{The gas mass profile \label{massprof}}
From the DIABOLO map, we can directly estimate the gas mass which gives rise
to the measured SZ signal. 
The SZ signal can directly converted into a gas mass value by 
applying the linear transformation:   
$M(\theta) = C \times F_{\nu}(\theta)$,
where C is a constant which depends on the redshift ($z=0.45$) and also on
the gas temperature. 
The SZ signal map is obtained after the subtraction of the point source best
fit model (see Sec.~\ref{szsignal}). This RXJ1347-1145 SZ map is shown on
figure~\ref{fig1}c (The X-ray contours have been overplotted). 
Afterwards, it is converted into a gas mass
map, which is then integrated with respect to the distance to the
cluster center (see Fig.~\ref{fig3}). 

Due to the size of the DIABOLO map and especially to the part used to
subtract the baseline (30\% on the edge of each line), we could  only measure
the projected gas mass within  a projected radius of 74'' (ie: 0.6Mpc): 
$M_{gas}^{SZ}(\theta<74'')=(1.1 \pm 0.1)\times 10^{14}\,
h_{50}^{-5/2}$M$_{\odot}$.  The error bar are quoted at a 68\% confidence
level.  This value  agrees very well with the estimate gas mass derived from
the X-ray best fit model (under the hypothesis of a spherical $\beta$-model): 
$M_{gas}^{\beta-model}(r<0.6\textrm{Mpc})= 10^{14}\, h^{-5/2}$M$_{\odot}$.  

This result confirms the agreement of the SZ and the X-ray data, when the
hypothesis of the $\beta$-model is adopted. In our case, no bias has been
introduced in the measured  projected mass by the baseline subtraction. In
fact the part of each line used to subtract the baseline has been taken out of
0.6Mpc from the cluster center. 
A radius of 0.6Mpc corresponds to an angular radius of 74.2''
(8.8$r_c$). Assuming a virial radius of 10$r_c$, the mass unaccounted for is
estimated as no more than 16\% of the total gas mass.

Our determination of the gas mass agrees with the X-ray determination.
However, the X-ray total mass determination disagree with the strong 
lensing \citep{sahu98} and the weak lensing \citep{fischer97}
determinations by a factor of two. 
\citet{allen98} explained this discrepancy by the effect of the cooling flow
on the total mass estimation through X-ray data.
He concludes that the correction from this effect removes the differences
between the estimators of the total mass. 
As an example, we can propose an increase of the gas temperature by a
factor of two ($\sim18$keV). To keep the SZ signal unchanged, this increase has
to be balanced by a decrease of the gas density by a factor of two. It then
affects our results in the following ways: 
(a) The infered total mass will be divided by a factor of two
($M_{gas}(r<0.6\textrm{Mpc})= (0.61 \pm 0.04)\times 10^{14}\,
h_{50}^{-5/2}$M$_{\odot}$).  
The total mass will also increase by a factor of two and then be in
agreement with the lensing measurements. 
(b) As far as we suggest an increase of the gas temperature balanced by a
decrease of the gas density, the SZ signal will remain unchanged.
On the opposite, the X-ray flux will decrease  by more than a factor of 2
(scaling the X-ray flux as $n_e(0)^2\, \sqrt(T_g)$). Consequently, the Hubble
constant will be affect by a factor of 2 ($H_0 \propto F_{SZ}^2/F_X$).

The previous scheme is very simple. The thermodynamic state of the intracluster
medium is probably more complex mainly made of two gas components. A cold
($\propto 9$keV), dense and peacked central component
surrounded by a hot ($\propto 18$keV), thin and  extended component.
In this case the cluster should present a non isothermal radial profile
(assuming the spherical symmetry). 
This kind of hypothesis could explain the very strong
degeneracy which we encountered while trying to determine the core radius (see
Sec.~\ref{szsignal}).  Unfortunately, the quality of our data is not good
enough to argue in this direction. 
Nevertheless, this suggestion is not unlikely. Recent XMM-Newton
observations have already shown such a behavior in other cooling flow
clusters: A1795 \citep{arnaud00,tamura01}, A1835 \citep{peterson00}

\subsection{Point sources \label{point}}

The best fit model gives a flux of $5.7\pm1.5$mJy/beam for the central point source.
This point source can be visualized on the data map after the subtraction of
the SZ component. Figure~\ref{fig1}d shows clearly an excess of
positive emission in its center. The NVSS map \citep{condon98} has been
overplotted as contours. Both signals match very well in term of position. 

This point source has already been observed at different frequencies.
It has been detected in the NVSS by \citet{condon98} with a flux
of $46.9\pm1.9$mJy at 1,4GHz. \citet{komatsu99} have reported
various measurements of this point source: an OVRO flux of $10.1\pm0.5$ at
28.5GHZ, a NMA (Nobeyama Millimeter Array) flux of $5\pm1.5$mJy at 100GHz and
a 2$\sigma$ upper limit of $4.8$mJy obtained with  SCUBA at 350GHz. 
From this set of measurements and assuming a power law spectrum for the point
source radio emission, they have derived:
$F_{RS}=(55.7\pm1.0)\nu_{GHz}^{-0.47\pm0.02}$mJy. 
More recently, \citet{komatsu00} have observed it with the VLA
at 8.46 and 22.46GHz. The respective detected flux are $22.42\pm0.04$ and
$11.5\pm0.17$mJy.

Using those latest measurements, we update the point source power spectrum
determination: $F_{RS}=(77.8\pm1.7)\nu_{GHz}^{-0.58\pm0.01}$~mJy. 
The quality of this fit is moderate, due to the very robust constraints given
by some of the data points (see Fig.~\ref{fig4}). 
It is difficult to conclude firmly that a single
power spectrum can model correctly the synchrotron emission from radio
wavelengths down to millimeter wavelengths. 
The  intensity and the spectral shape of the synchrotron emission is driven by
the internal magnetic field of the radio source. Furthermore its strength also
drives the maximum frequency of emission allowed.
To date, a lot of radio sources have been observed with various intensity and
spectral shapes. Because of this high rate of variation from one source to an
other and because of the lack of millimeter data, we are not able to state on
an eventual cut-off in the spectral shape of our radio source. For this
reason, we adopted the previously defined  power law model. It provides us an
estimation of the flux contributed by the central point source at 143GHz
(2.1mm). 

To explain the nature of the millimeter emission we have detected, we will now
compare the residual radio flux expected at 143GHz (2.1mm) and 350GHz
(850$\mu$m) from the previous power law spectrum  to the flux respectively 
obtained from the DIABOLO data and from the  SCUBA data. 
At 2.1mm, we have extrapolated:
$F_{RS}^{est}(\textrm{2.1mm}) = 4.3 \pm 0.3$mJy. This estimation is at
1$\sigma$  compatible with our determination from the DIABOLO map. 
The 350GHz (eg: 805$\mu$m) flux can be estimated too:
$F_{RS}^{est}(\textrm{0.85mm}) = 2.7 \pm 0.4$mJy. 
Now, from the SCUBA data points published by \citet{komatsu99}, 
we have subtracted the positive SZ contribution,
computed from the cluster SZ spectrum and scaled by our $y$ best fit value.
The SZ radial profile has been obtained from the model C (see
Sec.~\ref{szsignal}) convolved with the SCUBA beam (a Gaussian beam with
$\sigma_{FWHM}=15$''). We fixed the value of the signal offset (so-called DC
offset) to 2.7mJy/beam as published by Komatsu et al. We finally deduced for
the point source: $F_{RS}(\textrm{0.85mm})=1.8 \pm 1.0$mJy. 
The low signal to noise ratio of this last flux does not allow us to consider 
this result as the detection of a real submillimeter point source. 

Nevertheless, our DIABOLO flux and the SCUBA flux are both compatible with the
estimation extrapolated from the point source power spectrum. For this reason,
it seems reasonable to deduce  that the point source emission seen in our map
is likely due to the synchrotron millimeter tail of the central radio point
source emission.

\section{Conclusion \label{conclu}}

We have produced an extended map ($4''\times 4''$) of the RXJ1347-1145 cluster
with the DIABOLO photometer. 
We have drawn the distribution of the SZ signal up to 74''
(0.6Mpc) from the cluster center. This SZ signal is as much extended as the
X-ray emission and is the strongest SZ effect detected to date.

The SZ map allows us to directly derived the projected mass
distribution and by the way to measure the cluster gas mass up to 0.6Mpc.
We derived $M_{gas}=(1.1\pm0.1) \times 10^{14}\, h_{50}^{-5/2}$M$_{\odot}$. The gas mass
estimated within the same region, under the assumption of a classical
$\beta$-model and the cluster parameters derived from the X-ray data, agrees
with our value.  

Due to the map noise level and to the observing strategy, we are not able to
firmly conclude to the existence 
of substructure in the gas distribution. Nevertheless, our map presents some
asymmetric features at more than a 2$\sigma$ level.

At the moment, the first results of the Chandra and the XMM-Newton satellites, 
concerning the observation of galaxy clusters, already show some
departure in the density distribution from the $\beta$-model symmetry, as well
as in the temperature distribution from the isothermality
\citep{fabian00,vikhlinin00, arnaud00,tamura01,peterson00}.
For this reason, the comparison of the upcoming Chandra
and XMM-Newton X-ray obervations to the actual and the future SZ data (from
interferometers and bolometer arrays) are needed to understand the cluster
physical structure.

\acknowledgments

The authors want to thanks the IRAM staff (from astronomers to
cookers). Thanks to N. Aghanim, E. Komatsu, M. Hattori and Y. Suto for their
fruitful comments and discussions. 
We are very grateful to the anonymous referee for the quality of his report
that helped us to improved and clarify our paper. 
DIABOLO  is supported by the Programme National de Cosmologie,
Institut National pour les  Sciences de l'Univers, Minist\`ere de l'Education
Nationale de l'Enseignement Sup\'erieur et de la Recherche, CESR, CRTBT,
IAS-Orsay and LAOG.


\clearpage

\figcaption[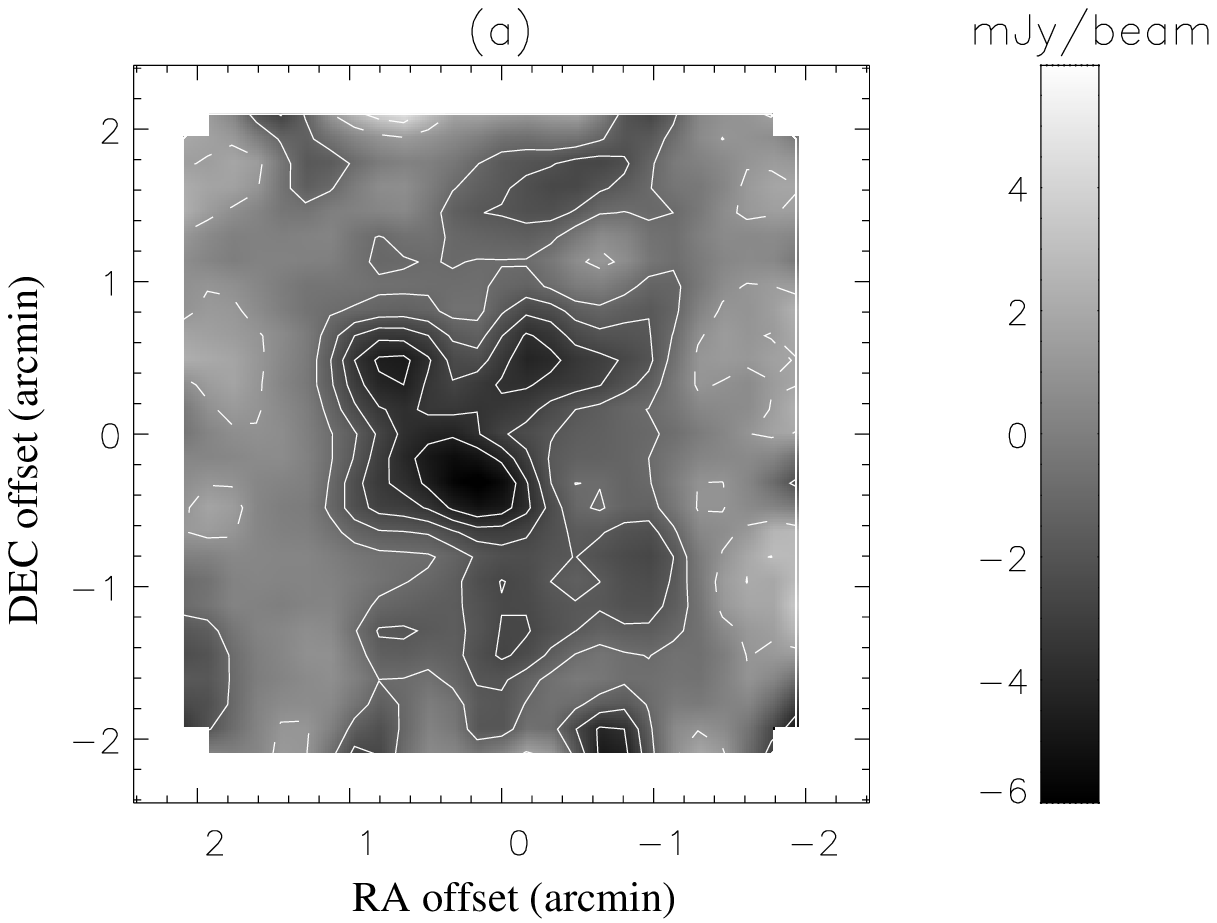,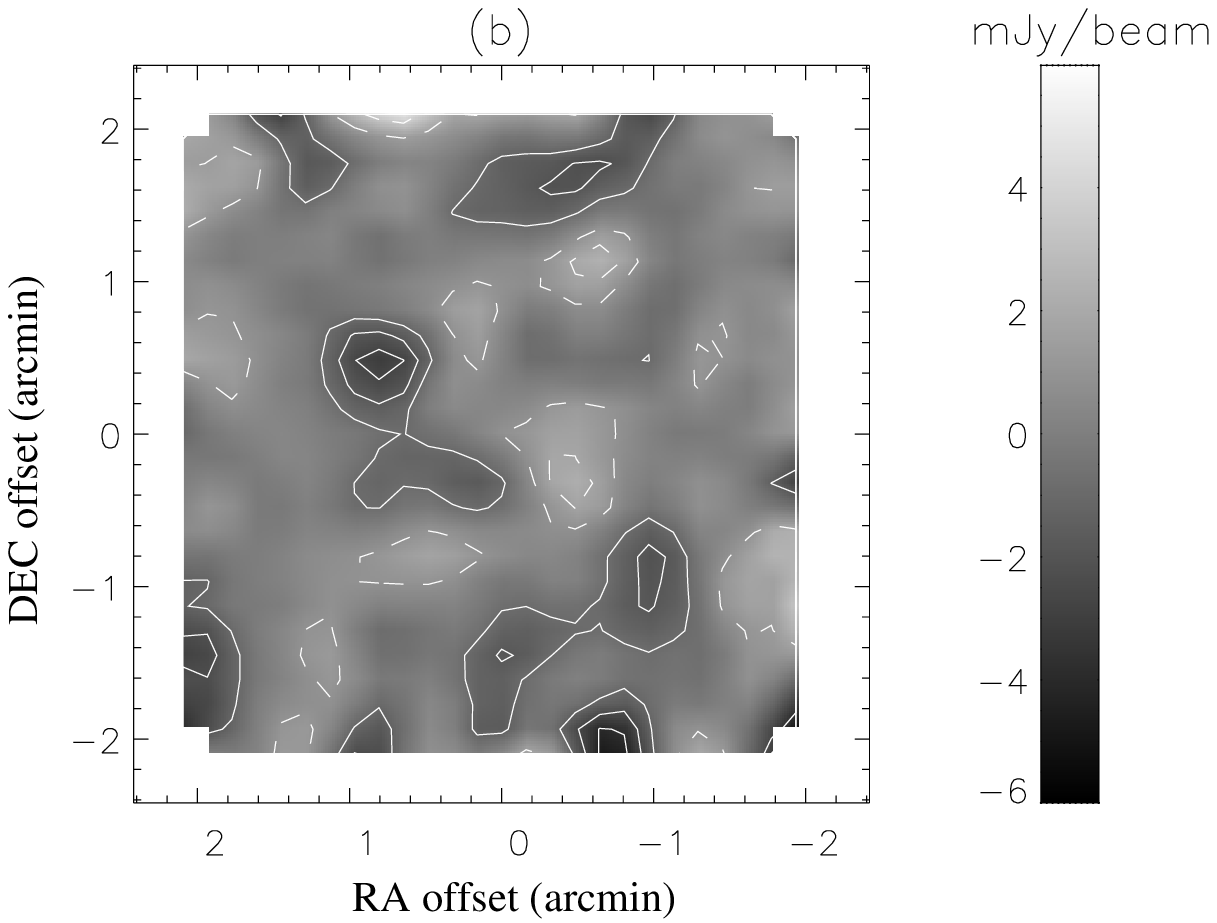,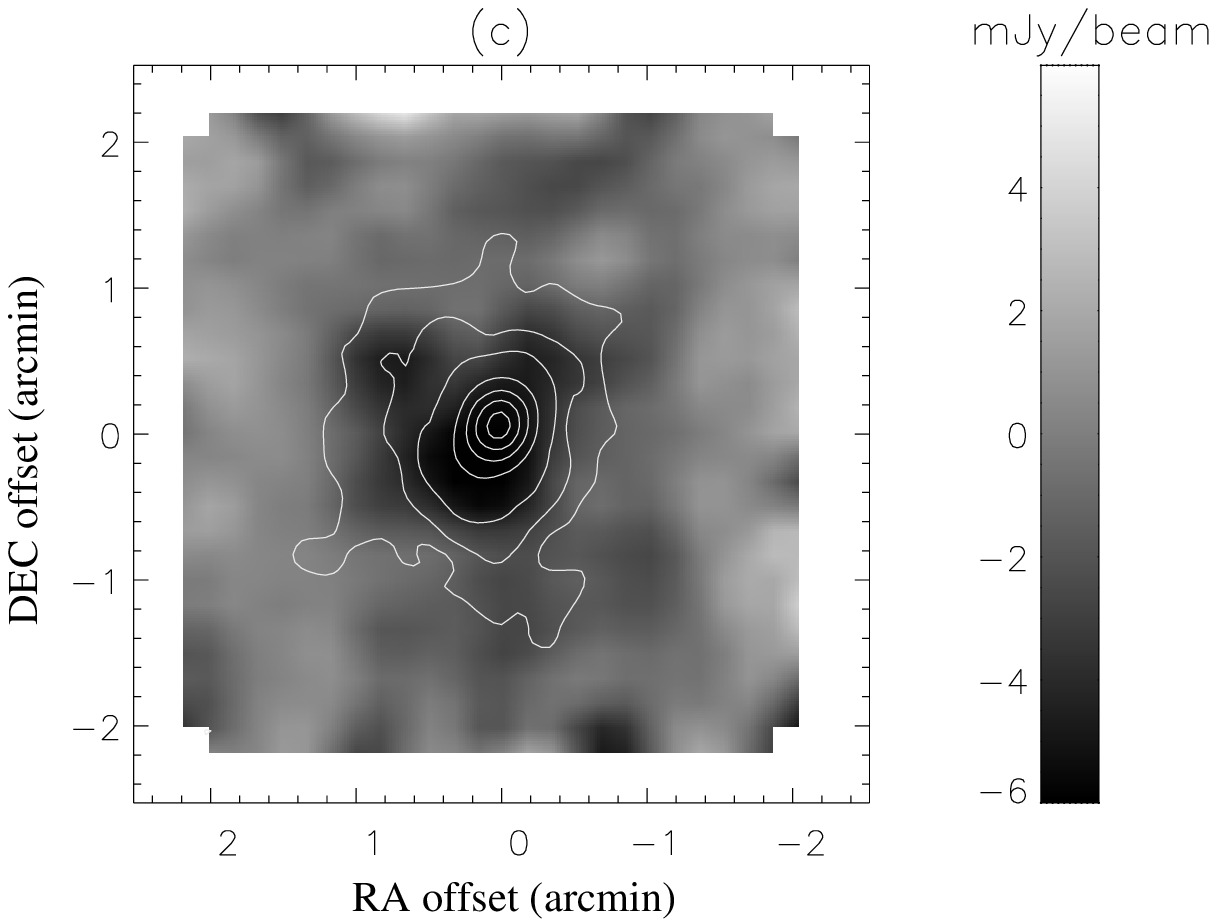,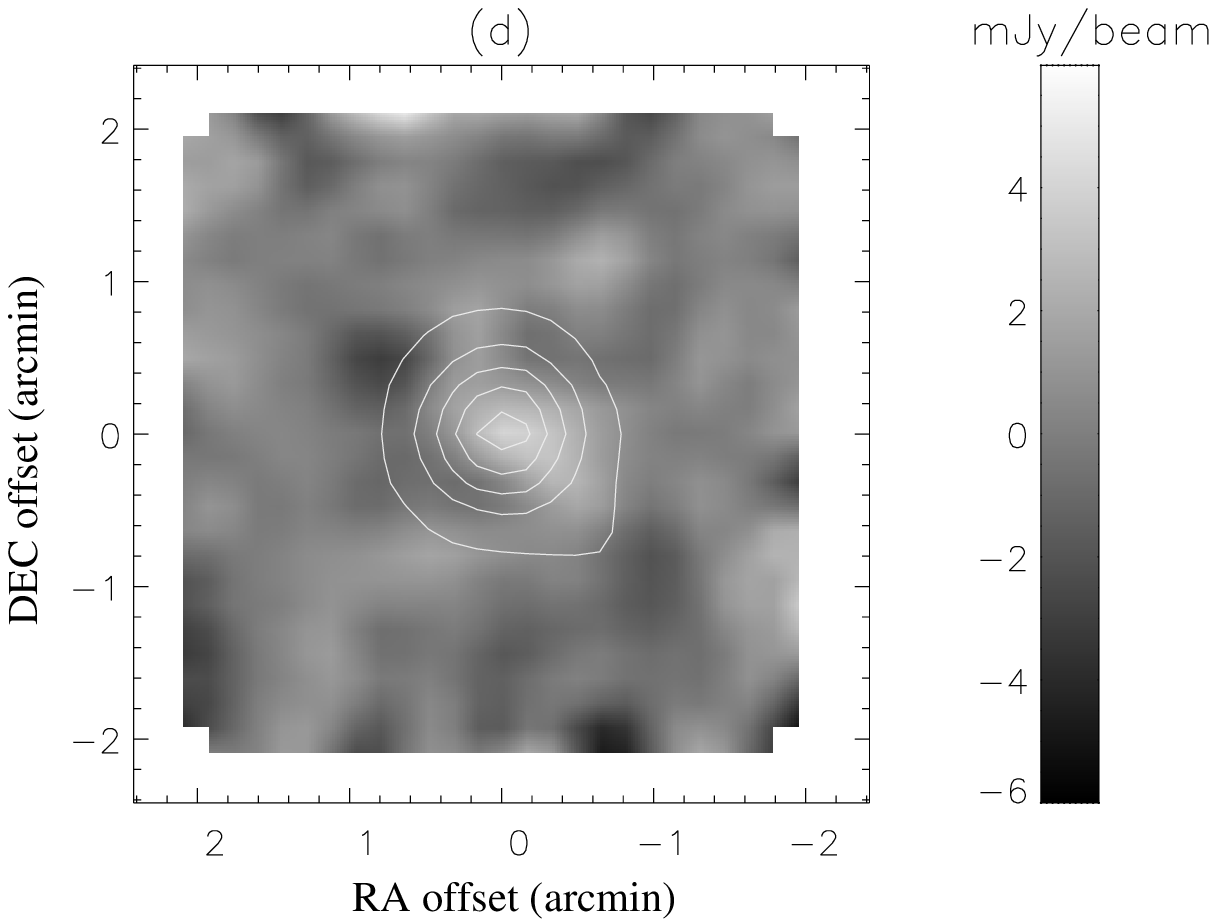]{ (a) 2.1mm map of RXJ1347-1145
  obtained with the DIABOLO 
  photometer. This map and the three others have been smoothed by a gaussian
  filter with 20'' FWHM.  
  The 1$\sigma$ noise level in term of flux is 1mJy in a pixel of
  $10''\times10''$  (equivalent to 0.5mJy/beam in the 22' DIABOLO beam and to
  0.3mJy/beam in the 30'' FWHM effective beam after smoothing). (b) Residual
  map after subtraction of the best fit model (SZ plus 
  point source, see text). In the maps {\it a} and {\it b} the overlying
  contours correspond to the 1, 2, 3, 4 and 5 $\sigma$ detection levels. The
  negative and positive contours are respectively drawn with solid and dashed
  lines. 
  (c) RXJ1347-1145 SZ signal (after subtraction of the point source best fit
  model) compared to the X-ray cluster emission. The different contours
  overplotted correspond to 3, 5, 10, 30, 50, 70 and 90\% of the X-ray maximum
  of emission. (d) Point source detection at
  2.1mm after subtraction of the SZ best fit model. The contours overlying
  correspond to the NVSS map (1, 10, 30, 50, 70 and 90\% of the radio maximum
  emission).  The central position in the four map correspond to the position
  of the X-ray center: $\alpha_{2000}=13^{\rm{h}}47^{\rm{m}}31^{\rm{s}}$,
  $\delta_{2000}=-11^{\circ}45'11"$. 
\label{fig1}}

\figcaption[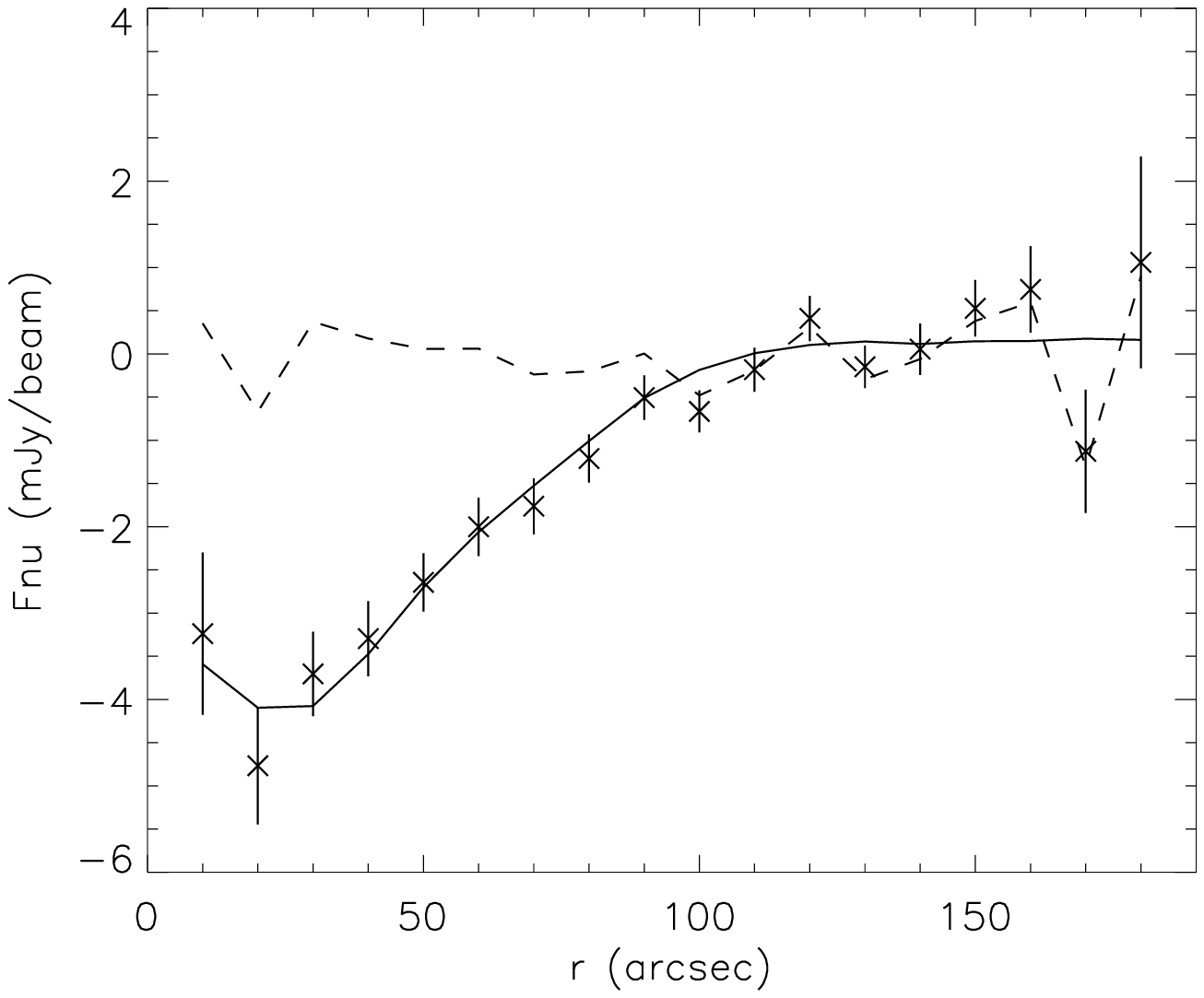]{RXJ1347-1145 radial profile at
  2.1mm computed from the DIABOLO 2.1mm map. 
  The data point are plotted with their associated 1$\sigma$ error bars. 
  The best fit model (including an SZ component and a point source component,
  see text) is overplotted (solid line). The residual radial profile is plotted
  as a dashed line. 
\label{fig2}}

\figcaption[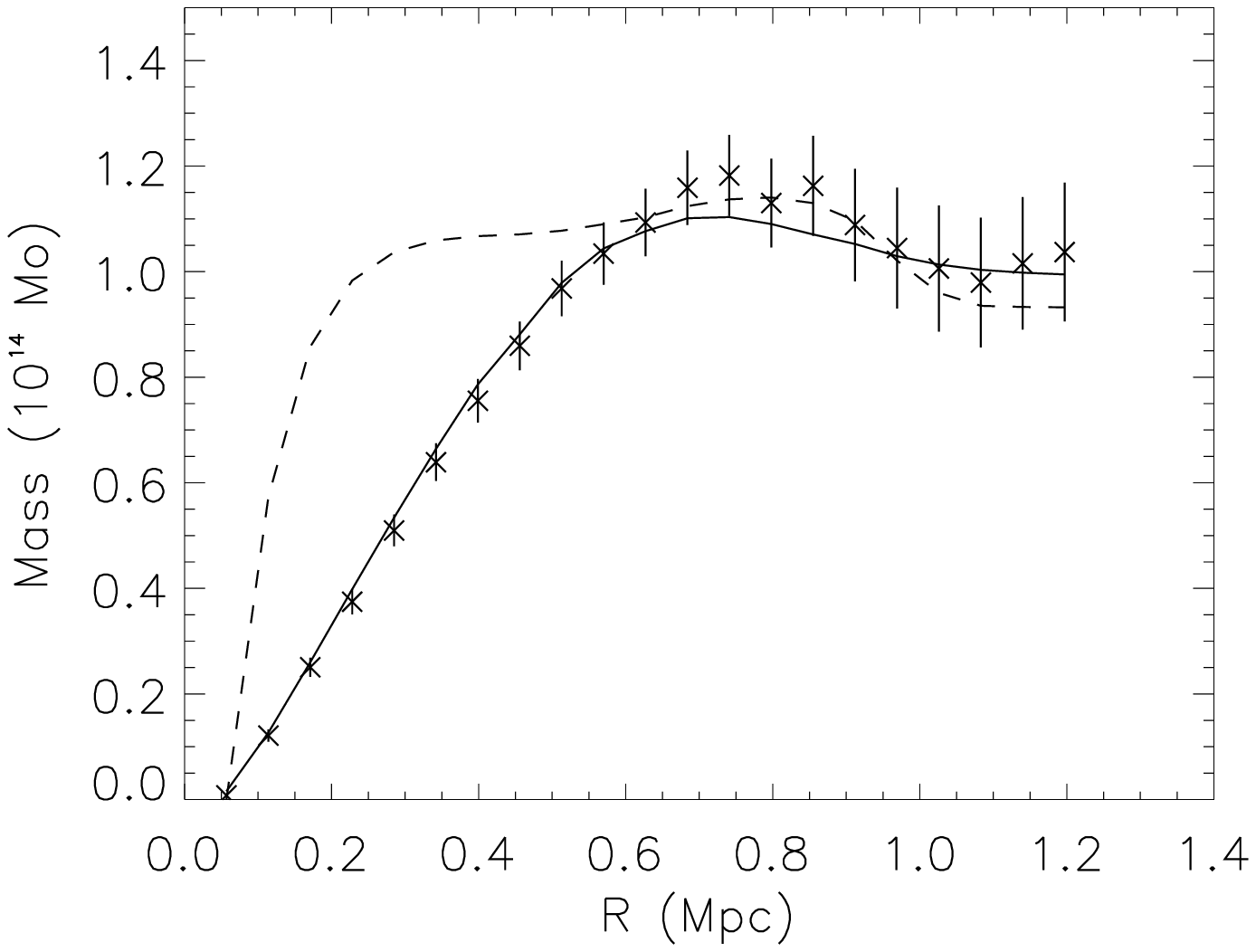]{RXJ1347-1145 gas mass profile computed directly from the
  2.1mm map. The solid line represents the gas mass profile of the SZ best fit
  model. The integral of the DIABOLO beam (dashed line) has been overplotted
  and scalled to the gas mass profile to show the extension of the SZ signal. 
\label{fig3}}

\figcaption[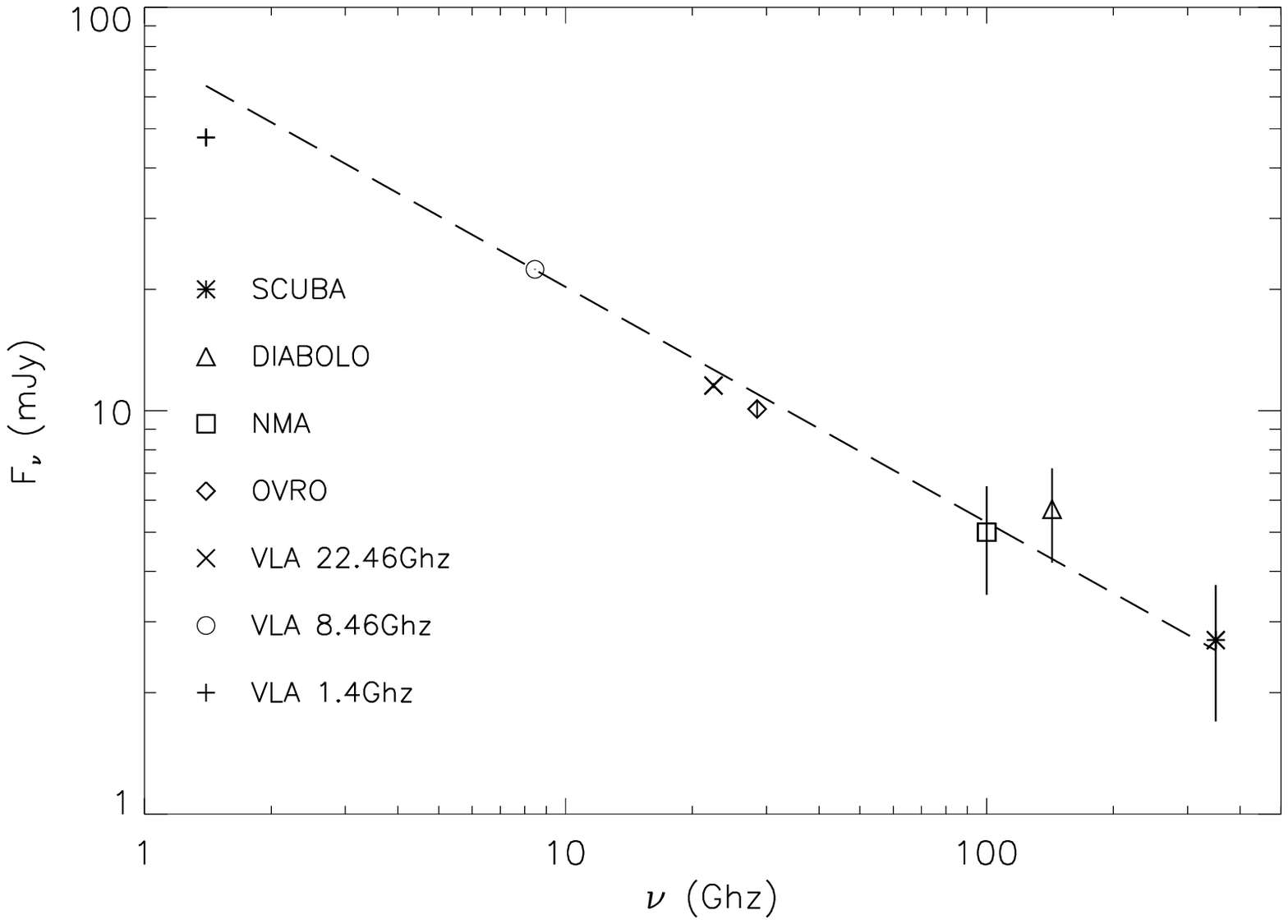]{Central point source spectum from 1.4 to 350GHz.  
\label{fig4}}

\clearpage

\begin{table}
\begin{center}
\caption{Best fit parameters. \label{table1}}
\begin{tabular}{cccccc}
\tableline\tableline
Models & $y(0)\; [10^{-4}]$   & $\theta_c$ (arcsec) & $\beta$ & $F_{RS}$ (mJy)&
$\chi^2/(n-1)$\tablenotemark{a} \\ 
\tableline
\textrm{A} & $4.1^{+1.7}_{-0.4}$ & $56.3^{+12.0}_{-19.1}$ 
& $0.89^{+0.26}_{-0.61}$ & 0\tablenotemark{b} & 1.1 \\ 

\textrm{B} & $5.3^{+0.8}_{-0.3}$ & $32.4^{+18.4}_{-21.5}$ 
& 0.56\tablenotemark{b} & 0\tablenotemark{b} & 0.9 \\ 

\textrm{C} & $5.7^{+0.3}_{-1.6}$ & $12.4^{+40.2}_{-12.4}$ 
&  $0.51^{+0.34}_{-0.11}$ & $3.4^{+2.6}_{-3.7}$ & 1.1 \\ 

\textrm{D} & $5.7^{+0.4}_{-0.6}$ & $14.8^{+28.7}_{-4.7}$
& 0.56\tablenotemark{b} & $4.1^{+39.3}_{-5.9}$ & 0.9 \\

\textrm{E} & $7.9^{+0.4}_{-0.5}$ & 8.4\tablenotemark{b} 
& 0.56\tablenotemark{b} & $5.7^{+1.4}_{-1.6}$ & 1.1 \\
\tableline
\end{tabular}
\tablenotetext{a}{where (n-1) is the number of degrees of freedom}
\tablenotetext{b}{fixed parameter, see text}
\end{center}
\end{table}

\clearpage

\begin{table}[!t]
\begin{center}
\caption{Fluxes in $50''\times 50''$  SE, NE, NW and SW regions as defined by
  \citet{komatsu00}. The column respectively gather the DIABOLO fluxes (1),
the residual flux after the subtraction  of the best fit model (2), the 
NOBA/NRO fluxes (3). Columns (1) and (3) have to be compared. The second 
column try to highlight some eventual excess (positive or negative). For each
collumn the associated 1$\sigma$ noise flux is quoted.
  \label{table2}} 
\begin{tabular}{cccc}
\tableline\tableline
Regions  & map  & residual map & NOBA/NRO \\ 
& (mJy) & (mJy) & (mJy)\\
\tableline
SE & $-12.2$ & $-0.8$ & $-11.3$ \\
NE & $-9.9$ & $-0.7$ & $-4.7$ \\
NW & $-10.1$ & $-0.4$ & $-3.3$ \\
SW & $-6.8$ & $\;2.8$ & $-6.1$ \\
1$\sigma$ noise & $1.4$ & $1.4$ & $2.0$ \\
\tableline
\end{tabular}
\end{center}
\end{table}

\clearpage
\includegraphics[width=7.5cm]{f1a.eps}%
\includegraphics[width=7.5cm]{f1b.eps}\\
\includegraphics[width=7.5cm]{f1c.eps}%
\includegraphics[width=7.5cm]{f1d.eps}

\clearpage
\includegraphics{f2.eps}

\clearpage
\includegraphics{f3.eps}

\clearpage
\includegraphics{f4.eps}

\end{document}